\documentstyle[eqsecnum,prb,preprint,aps,epsfig]{revtex}

\tightenlines

\newcommand{\be}{\begin{equation}}
\newcommand{\ee}{\end{equation}}
\newcommand{\bea}{\begin{eqnarray}}
\newcommand{\eea}{\end{eqnarray}}
\newcommand{\beaa}{\begin{eqnarray}}
\newcommand{\eeaa}{\end{eqnarray}}
\newcommand{\ba}{\begin{array}}
\newcommand{\ea}{\end{array}}
\newcommand{\bit}{\begin{itemize}}
\newcommand{\eit}{\end{itemize}}
\newcommand{\ben}{\begin{enumerate}}
\newcommand{\een}{\end{enumerate}}

\def\lab{\label}

\def\pa{\partial}

\def\al{\alpha}

\def\ga{\gamma}

\def\de{\delta}

\def\ep{\epsilon}

\def\la{\lambda}

\def\si{\sigma}

\def\om{\omega}
\def\Om{\Omega}
\begin{document}


\title{The r\^ole of the electromagnetic field in the formation of
domains in the process of symmetry breaking phase transitions}

\author{Emilio Del Giudice${}^{a}$ and Giuseppe Vitiello${}^{b}$
 \vspace{3mm}}

\address{${}^{a}$ INFN Sezione di Milano and Dipartimento di Fisica,
Universit\`a di Milano, I-20133 Milano, Italia
\\ [2mm] ${}^{b}$
Dipartimento di Fisica and INFN, Universit\`a di Salerno, I-84100
Salerno, Italy \vspace{2mm}}

\date{{\bf Version}, \today}

\maketitle

\begin{abstract}
In the framework of quantum field theory we discuss the emergence
of a  phase locking among the electromagnetic modes and the matter
components on an extended space-time region. We discuss the
formation of extended domains exhibiting  in their fundamental
states non-vanishing order parameters, whose existence is not
included in the Lagrangian. Our discussion is motivated by the
interest in the study of the general problem of the stability of
mesoscopic and macroscopic complex systems arising from
fluctuating quantum components in connection with the problem of
defect formation during the process of non-equilibrium symmetry
breaking phase transitions characterized by an order parameter.
\end{abstract}

\vspace{8mm}

P.A.C.S.: 11.10.-z, 64.90.+b, 11.30.Qc

\vspace{8mm}

\section{Introduction}

Complex systems made up of quantum components are usually
remarkably stable at mesoscopic and macroscopic space-time scales.
On the other hand, quantum fluctuations are the dominant feature
at the microscopic scale of the quantum components. The necessity
of taking into account such a double feature is reflected in the
usual quantum field theory (QFT) prescription that the Lagrangian
of the complex system built upon the quantum fields should be
invariant under the local phase transformation of the quantum
component field $\psi({\bf x},t) \rightarrow \psi'({\bf x},t) =
\exp(ig\theta({\bf x},t))\psi({\bf x},t)$. Local phase invariance
is the QFT solution to the problem of building a stable system out
of fluctuating components. The requirement of local phase
invariance demands the introduction of gauge fields, e.g. the
electromagnetic (e.m.) field $A_{\mu}({\bf x},t)$, such that the
Lagrangian be also invariant under the local gauge transformation
$A_{\mu}({\bf x},t) \rightarrow A_{\mu}'({\bf x},t) -
\partial_{\mu}\theta({\bf x},t)$. Such a transformation is devised
to compensate terms proportional to $\partial_{\mu}\theta({\bf
x},t)$ arising in the Lagrangian from the kinetic term for the
matter field $\psi({\bf x},t)$. This is a well known story. In the
present paper, given the above connection between the matter field
and the e.m. field,  we wish to discuss, in the frame of QFT, the
r\^ole played by the e.m. field in the locking of the phases of
the e.m. modes and of the matter components on an extended
space-time region. Furthermore, we will discuss extended domains
exhibiting in their fundamental states non-vanishing order
parameters, whose existence is not included in the Lagrangian.

The interest in the general problem of the stability of mesoscopic
and macroscopic complex systems arising from fluctuating quantum
components also finds one strong motivation in the study of the
physically relevant problem of defect formation during the process
of non-equilibrium symmetry breaking phase transitions
characterized by an order parameter \cite{Bunkov}. A topological
defect may indeed appear in such a process whenever a region,
surrounded by ordered domains, remains trapped in the "normal" or
symmetric state. Examples of topological defects are vortices in
superconductors and superfluids, magnetic domain walls in
ferromagnets, and many other extended objects in condensed matter
physics. On the other hand, topological defects, such as cosmic
strings in cosmology,  may have been also playing a r\^ole in the
phase transition processes in the early Universe \cite{kib}. The
phenomenological understanding of the defect formation in phase
transitions is provided by the Kibble-Zurek scenario
\cite{kib2,zurek1}. By considering the surprising analogy between
defect formation in solid state physics and in high energy physics
and cosmology \cite{volovik1}, it has been also stressed that the
analysis of the formation of defects in phase transitions becomes
a "diagnostic tool" \cite{zurek} in the study of non-equilibrium
symmetry breaking processes in a wide range of energy scales.
Questions such as why extended objects with topological
singularities are observed only in systems showing some sort of
ordered patterns, why defect formation is observed during the
processes of phase transitions, why  the features of the defect
formation are shared by quite different systems, from condensed
matter to cosmology, etc., have been specifically addressed in
refs. \cite{Alfinito:2001aa,Alfinito:2001mm} and the dynamics of
defect formation has been extensively studied in a large body of
literature in QFT; see, as general refs, \cite{Bunkov} and
\cite{Crakow}.

In these studies, in dealing with the presence of a gauge field in
the process of spontaneous symmetry breakdown a crucial r\^ole is
played by the well known Anderson-Higgs-Kibble (AHK) mechanism
\cite{hig,ItkZuber}, where the gauge field is expelled out of the
ordered domains and confined, through self-focusing propagation,
into "normal" regions, such as the vortex core, having a vanishing
order parameter, i.e. where the long range correlation modes (the
Nambu-Goldstone modes) responsible for the ordering are damped away.
In the present paper, going beyond the well established AHK
mechanism, our attention is focused on the dynamics governing the
radiative gauge field and, as said above, its r\^ole in the onset of
phase locking among the e.m. modes and the matter components. In the
AHK mechanism the gauge field removes the order in the regions where
it penetrates, thus describing the self-focusing gauge field
propagation in ordered condensed matter as well as in asymmetric
vacuum in elementary particle physics. Here we study the r\^ole of
radiative gauge field in sustaining the phase locking in the
coherent regime.

We choose as our model system an ensemble of a given number $N$ of
two-level atoms, which may represent rigid rotators endowed with an
electric dipole. The interaction of these atoms with the e.m.
quantum radiative modes will be considered. Moreover, we will
examine the effects on the system of an electric field generated by
an external source or else by an impurity introduced in the system,
thus making contact with the family of the so called Jaynes-Cummings
models \cite{JC}, extensively studied in the literature in
connection with quantum optics problems (see e.g. ref. \cite{Knight}
for detailed analysis). Our discussion in the present paper may be
indeed of some relevance to quantum optics, as well. As a matter of
fact, a system of $N$ two-level atoms interacting with quantized
e.m. modes has been known since long time to provide from a formal
point of view, under some convenient approximations and
restrictions, a strong analogy between the laser phase transition
and the conventional phase transitions in spontaneously broken
symmetry theories \cite{HakenVarenna,HakenBook}, although the
meaning of the constants in the potential function for the order
parameter is different. The key point in such a connection between
coherent (laser) light, the $N$ atom system and phase transition is
in the observation that, under convenient conditions, the behavior
of the e.m. mode is described by the potential
\be \lab{1} V(u, u^{*}) = - \al |u|^{2} + \beta |u|^{4} +
\frac{\al^{2}}{4 \beta} = \beta (|u|^{2} -
\frac{\al}{2\beta})^{2}~, \ee
where, in the Haken notation (see eq. (2.19) of
\cite{HakenVarenna} or eq. (VI.4.24) of  \cite{HakenBook}) $\al$
and $\beta$, with $\beta > 0$,  are convenient coefficients and
$u$ denotes the classical e.m. amplitude corresponding to the
quantum e.m. field amplitude (in the interaction representation
the electric field strength is decomposed as $E = u \exp (-i
\om_{0} t) + u^{*} \exp (i \om_{0} t)$ where $u$ is the slowly
varying amplitude and $\om_{0}$ the atomic resonant frequency).
The essential point is that the (mean value of the) "order
parameter" $u$ minimizing the potential $V(u,u^{*})$ is zero
(disordered or symmetric state) for $\al < 0$ and non-zero for
$\al > 0$, with $|u|^{2} = \frac{\al}{2\beta} \neq 0$ (ordered or
asymmetric state). In this latter case the system is said to be
above threshold (the threshold is set at $\al = 0$), i.e. it is
lasering. Of course, in the Lagrangian formalism the coefficient
$(- \al)$ denotes the "squared mass" of the field, whose sign, as
well known, controls the occurrence or not of spontaneous symmetry
breakdown. In the Haken interpretation $\al$ is the pump parameter
whose tuning may carry the system far from the equilibrium, i.e.
in the lasering region. Thus in the phase transition between the
disordered and the ordered state the order parameter $u$ changes
in time from zero to a value proportional to $\sqrt{\al}$. In his
analysis, Haken also considers the Hamiltonian in the interaction
representation
\be \lab{2} H = \hbar \ga (b^{\dag} S^{-} + ~b S^{+})~, \ee
which is a Jaynes-Cummings-like Hamiltonian, indeed. In eq.
(\ref{2}) $\ga$ is a coupling constant which is proportional to
the atomic dipole moment matrix element and to the inverse of the
volume square root $V^{-1/2}$, $b$ is the e.m. quantum field
operator (associated to the c-number amplitude $u$), $S^{\pm}$ are
the atomic polarization operators. In the Haken discussion the
atomic variables are integrated out at some point of the
computation since his interest is mostly focused on the e.m.
lasering effect. In our following analysis, instead, we keep them
and show that the phase locking between them and the e.m. mode can
be reached under convenient boundary conditions.


\section{The model}

Let us start by assuming that transitions between the atomic
levels are radiative dipole transitions. We thus disregard the
static dipole-dipole interaction. Moreover, the system is assumed
to be in a thermal bath kept at a non-vanishing temperature $T$.
Under such conditions the system is invariant under dipole
rotations. We use natural units $\hbar = 1 = c$. We assume the
system be spatially homogeneous and denote by $N$ the number of
atoms per unit volume. The $N$ atom system may be collectively
described by the complex dipole wave field $\phi ({\bf x},t)$. In
Section IV we will also use the known formal equivalence (see e.g.
Section III.6 of ref. \cite{HakenBook}) of the system of two-level
atoms with a system of $\frac{1}{2}$ spins. The dipole wave field
$\phi ({\bf x},t)$ integrated over the sphere of unit radius $\bf
r$ gives:
\be \lab{3} \int d\Om |\phi ({\bf x},t)|^{2} = N~, \ee
where $d\Om= \sin \theta d \theta d \phi$ is the element of solid
angle and $(r, \theta, \phi)$ are the polar coordinates of $\bf
r$. By introducing the rescaled field $\chi ({\bf x},t) =
\frac{1}{\sqrt{N}} \phi ({\bf x},t)$ eq. (\ref{3}) becomes
\be \lab{4}
 \int d\Om |\chi ({\bf x},t)|^{2} = 1~.  \ee

Since the atom density is assumed to be spatially uniform, the
only relevant variables are the angular ones. Thus, in full
generality, we may expand the field $\chi ({\bf x},t)$ in the unit
sphere in terms of spherical harmonics:
\be \lab{4a} \chi ({\bf x},t) = \sum_{l,m}
\al_{l,m}(t)Y^{m}_{l}(\theta, \phi) ~, \ee
which, by setting $\al_{l,m}(t) = 0$ for $l \neq 0,~ 1$, reduces
to the expansion in the four levels  $(l,m) = (0,0)$ and $(1,m), m
= 0, \pm 1$. The populations of  these levels are given by $N
|\al_{l,m}(t)|^{2}$ and at thermal equilibrium, in the absence of
interaction, they follow the Boltzmann distribution. Moreover, the
dipole rotational invariance implies that there is no preferred
direction in the dipole orientation, which means that the
amplitude of $\al_{1,m}(t)$ does not depend on $m$, and that no
permanent polarization may develop for such a system in such
conditions, i.e. the time average of the polarization $P_{{\bf
n}}$  along any direction ${\bf n}$ must vanish. We thus write
\bea \nonumber \al_{0,0}(t) &\equiv& a_{0}(t) \equiv
A_{0}(t)~e^{i\de_{0}(t)} ~,\\
 \al_{1,m}(t) &\equiv&  A_{1}(t)~e^{i\de_{1,m}(t)}~e^{-i\om_{0}t} \equiv a_{1,m}(t)
~e^{-i\om_{0}t} ~,\lab{9} \eea
where $ a_{1,m}(t) \equiv A_{1}(t)~e^{i\de_{1,m}(t)}$.
~$A_{0}(t)$, $A_{1}(t)$, $\de_{0}(t)$ and $\de_{1,m}(t)$ are real
quantities. In eqs. (\ref{9}) we have also used $\om_{0} \equiv
\frac{1}{I}$, where $I$ denotes the moment of inertia of the atom,
which gives a relevant scale for the system: $\om_{0} \equiv k =
\frac{2\pi}{\la}$ (note that the eigenvalue of $\frac{{\bf
L}^{2}}{2I}$ on the state $(1,m)$, ${\bf L}^{2}$ being the squared
angular momentum operator, is $\frac{l(l+1)}{2I} = \frac{1}{I} =
\om_{0}$). By setting the ${\bf z}$ axis parallel to ${\bf n}$ and
using the explicit expressions for the spherical harmonics
\bea \nonumber Y^{0}_{0} &=& \frac{1}{\sqrt{4\pi}}~,~~~ Y^{0}_{1}
=
\sqrt{\frac{3}{4\pi}}~ \cos~\theta~,\\
Y^{1}_{1} &=& - \sqrt{\frac{3}{8\pi}}~ \sin~\theta ~e^{i\phi} ~=~
- [Y^{-1}_{1}]^{*} ~, \lab{7}
 \eea
we find
\bea \nonumber P_{{\bf n}} &=& \int d\Om \chi^{*} ({\bf x},t)({\bf
x} \cdot {\bf n})\chi ({\bf x},t)\\ &=&
\frac{2}{\sqrt{3}}A_{0}(t)A_{1}(t)\cos(\om -\om_{0})t ~, \lab{8}
\eea
where $\om t \equiv \de_{1,0}(t) - \de_{0}(t)$ and whose time
average is zero, as it should be. This confirms that the three
levels $(1,m)$, $m = 0, \pm 1$ are in the average equally
populated under normal conditions and that, as said above, we can
safely write $\sum_{m} ~|\al_{1,m}(t)|^{2} =  3 ~|a_{1}(t)|^{2}$.
On the other hand, the normalization condition (\ref{4}) gives, at
any time $t$,
\be \lab{4b} |\al_{0,0}(t)|^{2} + \sum_{m} ~|\al_{1,m}(t)|^{2} =
|a_{0}(t)|^{2} + 3 ~|a_{1}(t)|^{2}   = 1~,  \ee
By defining $Q$ as
\be \lab{4d} Q \equiv  |a_{0}(t)|^{2} + 3 ~|a_{1}(t)|^{2}
 ~, \ee
we thus see from eq. (\ref{4b}) that
\be \lab{4c} \frac{\pa}{\pa t}Q = 0~.\ee
i.e
\be \lab{4ac} \frac{\pa}{\pa t}{|a_{1}(t)|^{2}} = -
\frac{1}{3}~\frac{\pa}{\pa t}{|a_{0}(t)|^{2}} ~.\ee
Due to eq. (\ref{3}) (and the rescaling adopted for $\chi({\bf
x},t)$), eq. (\ref{4c}) expresses nothing but the conservation of
the total number $N$ of atoms; it also means that, as shown in eq.
(\ref{4ac}), due to the rotational invariance, the rate of change
of the population in each of the levels $(1,m)$, $m = 0, \pm 1$,
equally contributes, in the average,  to the rate of change in the
population of the level $(0,0)$, at each time $t$. Consistently
with eq. (\ref{4b}), in full generality we can set the initial
conditions at $t = 0$ as
\be \lab{4f} |a_{0}(0)|^{2} = \cos^{2} \theta_{0}~,~~
|a_{1}(0)|^{2} = \frac{1}{3}\sin^{2} \theta_{0}~,~~0 < \theta_{0}
< \frac{\pi}{2}~. \ee
We exclude the values zero and $\frac{\pi}{2}$ since they
correspond to the physically unrealistic conditions for the state
(0,0) of being completely filled or completely empty,
respectively. By properly tuning the parameter $\theta_{0}$ in its
range of definition one can adequately describe the physical
initial conditions. For example, $\theta_{0} = \frac{\pi}{3}$
describes the equipartition of the field modes of energy $E(k)$
among the four levels $(0,0)$ and $(1,m)$, $|a_{0}(0)|^{2} \simeq
|a_{1,m}(0)|^{2}, ~m =0, \pm 1$, as typically given by the
Boltzmann distribution when the temperature $T$ is high enough,
$k_{B} T \gg E(k)$. As we will see, however, the lower bound for
the parameter $\theta_{0}$ is imposed by the dynamics in a
self-consistent way.

The possibility of obtaining a non-zero permanent polarization,
and thus the dipole ordering in the system, is crucially
conditioned by the ratio between the populations in the atomic
levels. Namely, suppose that the atom system is under the
influence of an electric field ${\bf E}$ due, e.g.,  to  an
impurity, or to any other external agent. Assume ${\bf E}$ to be
parallel to the ${\bf z}$ axis. Then the term
\be \lab{4p} {\cal H} = - {\bf d} \cdot {\bf E} ~, \ee
where ${\bf d}$ is the electric dipole moment of the atom, will be
added to the system energy and will break the dipole rotational
symmetry. It will produce the mixing between the states
$Y^{0}_{0}$ and $Y^{0}_{1}$: $Y^{0}_{0} \rightarrow Y^{0}_{0} \cos
\tau + Y^{0}_{1} \sin \tau$ and $Y^{0}_{1} \rightarrow - Y^{0}_{0}
\sin \tau + Y^{0}_{1} \cos \tau$, with
\be \lab{4pa} \tan \tau = \frac{\om_{0} - \sqrt{\om_{0}^{2} + 4
{\cal H}^{2}}}{2{\cal H}} ~. \ee
Due to the mixing thus induced, the polarization $P_{{\bf n}}$ is
now given by
\bea \nonumber P_{{\bf n}} &=& \frac{1}{\sqrt{3}} (A_{0}^{2}-
A_{1}^{2})
\sin 2\tau \\
 &+&  \frac{2}{\sqrt{3}} A_{0}(t)A_{1}(t) \cos 2\tau ~ \cos( [~ \om -
\sqrt{\om_{0}^{2} + 4 {\cal H}^{2}}~]t)  ~, \lab{4pb} \eea
to be compared with eq. (\ref{8}) and whose time average is
non-zero: $\overline{P_{{\bf n}}} = \frac{1}{\sqrt{3}} (A_{0}^{2}-
A_{1}^{2}) \sin 2\tau$. The non-zero difference in the level
populations $(A_{0}^{2}- A_{1}^{2})$ is therefore crucial in
obtaining the non-zero polarization. We will  study under which
conditions such an occurrence can be realized.

In the following, we restrict ourselves to the resonant radiative
e.m. modes, i.e. those for which $k = \frac{2\pi}{\la} = \om_{0}$,
and we use the dipole approximation, i.e. we put $\exp(i{\bf
k}\cdot {\bf x}) \approx 1$ in our formulas, since we are
interested in the macroscopic behavior of the system. This means
that the wavelengths of the e.m. modes we consider, of the order
of $\frac{2\pi}{\om_{0}}$, are larger (or comparable) than the
system linear size. Let $c_{r}({\bf k}, t)$ denote the radiative
e.m. field operator with polarization $r$ and $u_{r}({\bf k}, t) =
\frac{1}{\sqrt{N}} c_{r}({\bf k}, t)$ the rescaled one. The field
equations for our system are \cite{Knight,Heitler}:
\bea \nonumber i \frac{\partial \chi ({\bf x},t)}{\partial t} &=&
\frac{{\bf L}^{2}}{2I} ~\chi ({\bf x},t) - i \sum_{{\bf k}, r}
d~\sqrt{\rho }~\sqrt{\frac{k}{2}} ~ ({\bf \ep}_{r} \cdot {\bf
x})\\
 \nonumber &~&[u_{r}({\bf k}, t)~ e^{-ikt}
- u_{r}^{\dag}({\bf k}, t)~e^{ikt}]~~\chi ({\bf x},t) ~,\\
\lab{10} i \frac{\partial u_{r}({\bf k}, t)}{\partial t} &=& i ~d
\sqrt{\rho}~\sqrt{\frac{k}{2}} ~ e^{ikt} \int d \Om ({\bf \ep}_{r}
\cdot {\bf x}) |\chi ({\bf x},t)|^{2}~, \eea
where $d$ is the magnitude of the electric dipole moment, $\rho
\equiv \frac{N}{V}$ and ${\bf \ep}_{r}$ is the polarization vector
of the e.m. mode (for which the transversality condition ${\bf k}
\cdot {\bf \ep}_{r} = 0$ is assumed to hold). We remark that the
enhancement by the factor $\sqrt{N}$ appearing in the coupling $~d
~\sqrt{\rho}$ in eqs. (\ref{10}) is due to the the rescaling of
the fields. We will comment more on this point later on.

\section{The field equations and the phase locking}

By resorting to the discussion of the eqs. (\ref{10}) presented in
ref. \cite{DelGiudice:1988wg}, our task is now to analyze the
implications of these  equations  with reference to the r\^ole
played by the e.m. modes in the onset of the phase locking between
them and the dipole field.

First, let us observe that use of eqs. (\ref{4a}) (and (\ref{9}))
in (\ref{10}) gives the set of coupled equations
\bea \lab{11a} \dot{a}_{0}(t) &=& \Om
\sum_{m}~u_{m}^{*}(t)~a_{1,m}(t)\\
\lab{11b} \dot{a}_{1,m}(t) &=& - \Om
~u_{m}(t)~a_{0}(t) \\
\lab{11c} \dot{u}_{m}(t) &=& 2~\Om~ a_{0}^{*}(t)~a_{1,m}(t) ~,
\eea
where $a_{1,m}(t) \equiv \al_{1,m}(t) ~e^{i\om_{0}t}$ (cf. eq.
(\ref{9})), the dot over the symbols denotes the time derivative,
$u_{m}$ is the amplitude of the e.m. mode coupled to the
transition $(1,m) \leftrightarrow (0,0)$ and $\Om \equiv
\frac{2d}{\sqrt{3}}\sqrt{\frac{\rho}{2\om_{0}}}~\om_{0} \equiv G
~\om_{0}$.

Eqs. (\ref{11a})-(\ref{11c}), as well as the eqs. (\ref{10}), from
which they have been derived, appear to be not invariant under
time-dependent phase transformations of the field amplitudes. We
want to investigate how gauge invariance can be recovered.

Use of the conjugate of eq. (\ref{11b}) in (\ref{11a}) gives,
consistently with eq. (\ref{4c}), the conservation law $\dot{Q} =
0$. Moreover, use of the conjugate of eq. (\ref{11b}) in
(\ref{11c}) leads to
\be \lab{14} \frac{\pa}{\pa t} {|u_{m}(t)|^{2}} = -
2~\frac{\pa}{\pa t} {|a_{1,m}(t)|^{2}}~.\ee
Since the amplitude $|\al_{1,m}(t)| = |a_{1,m}(t)|$ does not
depend on $m$, eq. (\ref{14}) shows that also the amplitude
$|u_{m}(t)|$ does not depend on $m$. Eq. (\ref{14}) shows the
existence of another constant of motion; namely, putting $|u(t)|
\equiv |u_{m}(t)|$ and using $|a_{1}(t)| \equiv |a_{1,m}(t)|$, we
can write
\be \lab{15} |u(t)|^{2} + 2~|a_{1}(t)|^{2} =
\frac{2}{3}\sin^{2}\theta_{0} ~,\ee
for any $t$, where we have also used the initial condition
(\ref{4f}) and set
\be \lab{4f2} |u(0)|^{2} = 0 ~.\ee
We observe that since $|u(t)|^{2} > 0$ eq. (\ref{15}) imposes
$|a_{1}(t)|^{2} \leq \frac{1}{3}\sin^{2}\theta_{0}$ and therefore
$|a_{0}(t)|^{2} \geq \cos^{2}\theta_{0}$ due to (\ref{4b}).  Note
that (\ref{15}) gives (cf. (\ref{4f}))
\be \lab{16} |u(t)|^{2} = 2(~|a_{1}(0)|^{2} - |a_{1}(t)|^{2}~)
~,\ee
for any $t$, which, by exploiting (\ref{4f}), reads
\be \lab{16a} |u(t)|^{2} = \frac{2}{3}~(~|a_{0}(t)|^{2} - \cos^{2}
\theta_{0}~) ~.\ee
It is also useful to write
\be \lab{22}   u_{m}(t) = U (t) e^{i\varphi_{m}(t)}~, \ee
with $U (t)$ and $\varphi_{m}(t)$ real quantities.

By combining eqs. (\ref{9}) and (\ref{22}) with
(\ref{11a})--(\ref{11c}) and equating real and imaginary parts, we
get
\bea \lab{24aa}
\dot{A_{0}}(t) &=& \Om U (t)A_{1}(t) \cos \al_{m} (t) ~,\\
\lab{24ab} \dot{A_{1}}(t) &=& - \Om U (t) A_{0}(t) \cos \al_{m} (t) ~,\\
\lab{24c} \dot{U}(t) &=& 2\Om A_{0}(t)A_{1}(t) \cos \al_{m} (t) ~,\\
\lab{24d} \dot{\varphi_{m}}(t) &=& 2\Om
\frac{A_{0}(t)A_{1}(t)}{U(t)}\sin \al_{m} (t) ~, \eea
where we have put
\be \lab{24e} \al_{m} \equiv \de_{1,m}(t) - \de_{0}(t) -
\varphi_{m}(t) ~. \ee
Similarly, we can derive equations for $\dot{\de_{1,m}}$ and
$\dot{\de_{0}}$.

From eqs. (\ref{24aa})--(\ref{24c}) we see that since their left
hand sides are independent of $m$, so  the right hand sides have to
be, i.e. either $\cos \al_{m} (t) = 0$ for any $m$ at any $t$, or
$\al_{m}$ is independent of $m$ at any $t$. In both cases, eq.
(\ref{24d}) shows that $\varphi_{m}$ is then independent of $m$,
which in turn implies, together with eq. (\ref{24e}), that
$\de_{1,m}(t)$ is independent of $m$. Phases thus turn out to be
independent of $m$. We will therefore put $\varphi \equiv
\varphi_{m}$, $\de_{1}(t) \equiv \de_{1,m}(t)$ and $\al \equiv
\al_{m}$. We observe that in general the phases can be always
changed by arbitrary constants. The  independence of m of the phases
is dictated by the requirement to not violate the gauge invariance.
Should exist a difference between the phases having different m,
this difference could be changed by a rotation of the axes and would
spoil the gauge invariance. In the present case, the independence of
$m$ of the phases is of dynamical origin and we will find that the
phase locking among $\de_{0}(t)$, $\de_{1}(t)$ and $\varphi (t)$,
has indeed the meaning of recovering the gauge invariance. We will
discuss this point in the subsection II.B.

Summarizing, we can now write  $u(t) \equiv u_{m}(t)$ and
$a_{1}(t) \equiv a_{1,m}(t)$ and from the  equations
(\ref{11a})-(\ref{11c}) we get the known \cite{DelGiudice:1988wg}
set of equations:
\bea \lab{17a} \dot{a}_{0}(t) &=& 3~\Om~u^{*}(t)~a_{1}(t)\\
\lab{17b} \dot{a}_{1}(t) &=& - \Om~
u(t)~a_{0}(t) \\
\lab{17c} \dot{u}(t) &=&  2~\Om~ a_{0}^{*}(t)~a_{1}(t) ~. \eea
Eqs. (\ref{17a})-(\ref{17c}) are fully consistent with the
original normalization condition (\ref{4}) (or (\ref{4b})), with
the conservation (\ref{4c}) and with the dipole rotational
invariance expressed by the zero average polarization (cf. eq.
(\ref{8})). In eq.(\ref{17a}) the rate of change of the amplitude
of the level $(0,0)$ is shown to depend on the coupling between
the levels $(1,m)$, $m = 0, \pm 1$ and the radiative e.m. mode of
corresponding polarization. Each of these couplings contribute in
equal measure, due to rotational invariance, to the transitions to
(0,0). Similarly, in eq.(\ref{17b}) the rate of change of the
amplitude of each level $(1,m)$ is shown to depend on the coupling
between the the level $(0,0)$ and the corresponding radiative e.m.
mode. Finally, in eq.(\ref{17c}) the transitions $(0,0)
\leftrightarrow (1,m), m = 0, \pm 1$ control the rate of change of
the amplitude of the radiative e.m. mode of corresponding
polarization. These equations thus reflect the correct selection
rules in radiative and absorption processes
\cite{Herzberg,Powell,UmezVit}. Eq. (\ref{17a}) describes the fact
that each of the levels $(1,m)$ may find in the e.m. field the
proper mode to couple with, in full respect of the selection
rules. In this sense, the field concept, as a full collection of
e.m. modes with all possible polarizations, is crucial here. As
already said, eqs. (\ref{17a})-(\ref{17c}) are fully  consistent
with the physical boundary conditions and the motion equation
(\ref{10}) from which they are derived.

\subsection{The ground state}

We want  to study now the vacuum or ground state of the system for
each of the modes $a_{0}(t)$, $a_{1}(t)$ and $u(t)$ described by
(\ref{17a})-(\ref{17c}) (i.e. by (\ref{10})).

It is convenient to differentiate once more with respect to time
both sides of (\ref{17a}). By using (\ref{17b}) and (\ref{17c})
and the constants of motion (\ref{4b}) and (\ref{15}) we eliminate
the  variables $a_{1}(t)$ and $u(t)$. We thus find:
\be \lab{27}  \ddot{a}_{0} (t)  =
4~{\Om}^{2}\ga_{0}^{2}(\theta_{0}) a_{0}(t) -
4~{\Om}^{2}|a_{0}(t)|^{2}a_{0}(t)~, \ee
where $\ga_{0}^{2}(\theta_{0}) \equiv \frac{1}{2}(1 + \cos^{2}
\theta_{0})$. Eq. (\ref{27}) can be written in the form
\be \lab{28}  \lab{21} \ddot{a}_{0} (t)  = - \frac{\de }{\de
a_{0}^{*}}V[a_{0}(t), a_{0}^{*}(t)]~,\ee
where the potential $V[a_{0}(t), a_{0}^{*}(t)]$ is
\be \lab{28b} V[a_{0}(t), a_{0}^{*}(t)] = 2 {\Om}^{2}
(|a_{0}(t)|^{2} - \ga_{0}^{2}(\theta_{0}))^{2}
 ~. \ee
In a standard fashion (see e.g. \cite{ItkZuber}) we may adopt the
semiclassical ('mean field') approximation in order to study the
ground state of the theory. We thus search for the minima of the
potential $V$. Let $a_{0,R}(t)$ and $a_{0,I}(t)$ denote the real
and the imaginary component, respectively, of the $a_{0}(t)$
field: $|a_{0}(t)|^{2} = A_{0}^{2}(t)= a_{0,R}^{2}(t) +
a_{0,I}^{2}(t)$.  The potential has a relative maximum at $a_{0} =
0$ and a (continuum) set of minima given by
\be \lab{29}  |a_{0}(t)|^{2} =  \frac{1}{2} (1 + \cos^{2}
\theta_{0}) = \ga_{0}^{2}(\theta_{0}) ~. \ee
These minima correspond to the points on the circle of squared
radius $\ga_{0}^{2}(\theta_{0})$ in the $(a_{0,R}(t),a_{0,I}(t))$
plane. We thus recognize that we are in the familiar case of a
theory where the cylindrical $SO(2)$ symmetry (the phase symmetry)
around an axis orthogonal to the plane $(a_{0,R}(t),a_{0,I}(t))$
is spontaneously broken. The order parameter is given by $\ga_{0}
(\theta_{0})$. Note that eq. (\ref{29}) does not fix the (ground
state expectation) value of the phase field $\de_{0}(t)$. The
points on the circle represent (infinitely many) possible vacua
for the system and they transform into each other under shifts of
the field $\de_{0}$: $\de_{0} \rightarrow \de_{0} + \al$ (SO(2)
rotations in the $(a_{0,R}(t),a_{0,I}(t))$ plane). The phase
symmetry is broken when one specific ground state is singled out
by fixing the value of the $\de_{0}$ field.

By proceeding as usual in these circumstances \cite{ItkZuber}, we
transform to new fields: $A_{0}(t) \rightarrow A_{0}'(t) \equiv
A_{0}(t) - \ga_{0}(\theta_{0})$ and $\de_{0}'(t) \rightarrow
\de_{0}(t)$, so that $A_{0}'(t)  = 0$ in the ground state for
which $A_{0}(t) = \ga_{0}(\theta_{0})$. Use of these new variables
in $V$ leads to recognize that the amplitude $A_{0}'(t)$ describes
a quasi-periodic mode with pulsation $m_{0}
 = 2{\Om} \sqrt{(1 + \cos^{2} \theta_{0})}$ (a 'massive' mode
with real mass $2{\Om} \sqrt{(1 + \cos^{2} \theta_{0})}$) and that
the field $\de_{0}'(t)$ corresponds to a zero-frequency mode (a
massless mode) playing the r\^ole of the so-called Nambu-Goldstone
(NG) field or collective mode implied by the spontaneous breakdown
of symmetry.

We note that when eq. (\ref{29}) holds, use of eqs. (\ref{4b}) and
(\ref{15}) gives  $A^{2}_{1} = \frac{1}{6} \sin^{2} \theta_{0}$ and
$\overline{U}^{2} = \frac{1}{3} \sin^{2} \theta_{0}$, moving away
from the initial condition values (\ref{4f}) and (\ref{4f2}),
respectively. In this respect, it is remarkable that the value
$a_{0} = 0$, which we have excluded in our initial conditions, cf.
eq. (\ref{4f}), on the basis of physical considerations,
consistently appears to be the relative maximum for the potential,
and therefore an instability point out of which the system
(spontaneously) runs away. Moreover, as already observed, use of the
constant of motion laws (\ref{4b}) and (\ref{15}) shows that
$|a_{0}|^{2} = 0$ implies $U^{2} = - \frac{2}{3}\cos^{2}\theta_{0}$
which is not possible since $U$ is real. Finally, we remark that the
bound  $|a_{0}(t)|^{2} \geq \cos^{2}\theta$ discussed above (see the
comment after eq. (\ref{4f2})) is consistently satisfied by
$|a_{0}(t)|^{2} = \ga_{0}^{2}(\theta_{0})$ (see eq. (\ref{29})).

We consider now the time derivative of both sides of (\ref{17b})
and by simple manipulations we find the following equation for the
amplitude $a_{1}(t)$:
\be \lab{31}  \ddot{a}_{1} (t)  =  - \si^{2} a_{1}(t) +
12~{\Om}^{2}|a_{1}(t)|^{2}a_{1}(t)~, \ee
where $\si^{2} = 2~{\Om}^{2}(1 + \sin^{2} \theta_{0})$. The
potential from which the r.h.s. of eq. (\ref{31}) is derivable is
\be \lab{32}   V[a_{1}(t), a_{1}^{*}(t)] = \si^{2}|a_{1}(t)|^{2} -
6 {\Om}^{2} (|a_{1}(t)|^{2})^{2} ~.\ee
In this case there is a relative minimum at $a_{1} = 0$ and a
(continuum) set of relative maxima on the circle of squared radius
\be \lab{32}  |a_{1}(t)|^{2}  = \frac{1}{6} (1 + \sin^{2}
\theta_{0}) \equiv \ga_{1}^{2}(\theta_{0})  ~. \ee
Note that, for $|a_{1}(t)|^{2} =  \ga_{1}^{2}(\theta_{0})$, $U^{2}
= - \frac{1}{3} \cos^{2} \theta_{0} < 0$, which is not acceptable
since  $U$ is real. The values on the circle of radius
$\ga_{1}(\theta_{0})$ are thus forbidden for the amplitude
$A_{1}$. This is consistent with the intrinsic instability of the
excited levels $(1,m)$. We have also seen that the conservation
law (\ref{15}) and the reality condition for $U$ require that
$|a_{1}(t)|^{2} \leq \frac{1}{3}\sin^{2}\theta_{0}$ which lies
indeed below $\ga_{1}^{2}(\theta_{0})$, and we note that the value
$\frac{1}{6} \sin^{2} \theta_{0}$ taken by $A^{2}_{1}$ when
$|a_{0}(t)|^{2} = \ga_{0}^{2}(\theta_{0})$ also lies below the
bound. In conclusion,  the potential $V[a_{1}(t), a_{1}^{*}(t)]$
involved in the dynamics must be lower than
$\frac{1}{3}\sin^{2}\theta$.

This is enough about the consistency between eqs. (\ref{27}) and
(\ref{31}). As mentioned above, we exclude that the amplitude
$A_{1}$ be constantly zero (at the minimum of $V[a_{1}(t),
a_{1}^{*}(t)]$), since this would correspond to the physically
unrealistic situation of the $(0,0)$ level completely filled. In
conclusion, within these dynamical bounds, the field $a_{1}(t)$
described by Eq. (\ref{31}) is  a massive field with (real) mass
(pulsation) $\si^{2} =2~{\Om}^{2}(1 + \sin^{2} \theta_{0})$.

Finally, we focus on the e.m. mode $u(t)$ and consider eq.
(\ref{17c}). By proceeding as above by differentiating  once more
with respect to time  we find
\be \lab{18}  \ddot{u} (t)  = - \mu^{2}u(t) -
6~{\Om}^{2}|u(t)|^{2}u(t)~, \ee
where $\mu^{2} = 2~{\Om}^{2}\cos 2\theta_{0}$. The r.h.s. of eq.
(\ref{18}) is derivable from the potential
\bea \nonumber  V[u(t), u^{*}(t)] &=& \mu^{2}|u(t)|^{2} +
3~{\Om}^{2}|u(t)|^{4} + ~\frac{1}{3} {\Om}^{2} \cos^{2}
2\theta_{0} \\
\lab{20} &=& 3 {\Om}^{2} (|u(t)|^{2} + \frac{1}{3} \cos
2\theta_{0})^{2}
~,\eea
and we note that $V[u(t), u^{*}(t)]$ is nothing but the potential
for the e.m. mode given in eq. (\ref{1}) for $- \al = \mu^{2}$ and
$\beta = 3~{\Om}^{2}$. We are in the case of a theory where the
symmetry can be spontaneously broken or not, according to  the
negative or positive value of the squared mass  $\mu^{2}$ of the
field (the pump in the Haken interpretation), respectively.

Again, in the semiclassical  approximation  we search for the minima
of the potential $V[u(t), u^{*}(t)]$ and see that $\mu^{2} \geq 0$
for $\theta_{0} \le \frac{\pi}{4}$ and the only minimum is at $u_{0}
= 0$. Eq. (\ref{18}) then describes quasi-periodic modes with
pulsation $\mu = {\Om} \sqrt{2~\cos 2\theta_{0}}$, typically
expected for a paraboloid potential $V[u(t), u^{*}(t)]$ with
cylindrical $SO(2)$ symmetry about an axis orthogonal to the plane
$(u_{R}(t),u_{I}(t))$ and minimum at $u_{0} = 0$. Here $u_{R}(t)$
and $u_{I}(t)$ denote the real and the imaginary component,
respectively, of the $u(t)$ field. In such a case we have the
symmetric solution with zero order parameter $u_{0} = 0$. This
solution describes the system when the initial condition, eq.
(\ref{4f2}), holds at any time. This occurrence is, however, not
consistent with the dynamical evolution of the system moving away
from the initial conditions exhibited by eq. (\ref{27}), as
mentioned above. Luckily, consistency is  dynamically recovered
provided $\theta_{0} > \frac{\pi}{4}$. In such a case, indeed,
$\mu^{2} = 2 {\Om}^{2} \cos 2\theta_{0} < 0$ and the potential has a
relative maximum at $u_{0} = 0$ and a (continuum) set of minima
given by
\be \lab{23}  |u(t)|^{2}  = - \frac{1}{3} \cos 2\theta_{0} = -
\frac{\mu^{2}}{6{\Om}^{2}} \equiv v^{2}(\theta_{0}) ~, ~~~~
 \theta_{0} > \frac{\pi}{4}. \ee
The  fact that in the present case $u_{0} =0$ is a maximum for the
potential means that the system dynamics evolves away from it,
consistently with the similar situation noticed above for the
$a_{0}$ mode where the system spontaneously evolves away from the
initial conditions. The symmetric solution of the minimum at
$u_{0} = 0$ is thus excluded for internal  consistency and the
lower bound $\frac{\pi}{4}$ for $\theta_{0}$ is thus dynamically
imposed in a self-consistent way.

In eq. (\ref{23}) the minima are the points of the circle of
squared radius $v^{2}(\theta_{0})$ in the $(u_{R}(t),u_{I}(t))$
plane. As in the case of the amplitude $a_{0}$ analyzed above, the
points on the circle represent (infinitely many) possible vacua
for the system and they transform into each other under shifts of
the field $\varphi$: $\varphi \rightarrow \varphi + \al$. For
$\theta_{0}
> \frac{\pi}{4}$ the phase symmetry is broken, the order parameter
is  given by $v(\theta_{0}) \neq 0$ and one specific ground state
is singled out by fixing the value of the $\varphi$ field.

As usual \cite{ItkZuber}, we transform to new fields: $U(t)
\rightarrow U'(t) \equiv U(t) - v(\theta_{0})$ and $\varphi'(t)
\rightarrow \varphi(t)$ so that in the ground state $U'(t) = 0$. Use
of these new variables in $V[u(t), u^{*}(t)]$ shows that the
amplitude $U'(t)$ describes a 'massive' mode with real mass
$\sqrt{2|\mu^{2}|} = 2\Om \sqrt{|\cos 2\theta_{0}|}$ (a
quasi-periodic mode) and that the field $\varphi'(t)$ is a
zero-frequency mode (a massless mode). This field, also called the
"phason" field \cite{Umez}, plays the r\^ole of the  Nambu-Goldstone
(NG) collective mode in the theories where symmetry is spontaneously
broken. When eq. (\ref{23}) holds, it is  $A_{1}^{2} = \frac{1}{6}$
which lies below the upper bound $\frac{1}{3}\sin^{2}\theta$
provided $\theta
> \frac{\pi}{4}$. Similarly, eq. (\ref{23}) implies  $A_{0}^{2} =
\frac{1}{2}$ which satisfies the constraint of being greater than
$\cos^{2}\theta$ for $\theta > \frac{\pi}{4}$.

In conclusion, the e.m. field, as an effect of the spontaneous
breakdown of the phase symmetry (eq. (\ref{23}))  gets a massive
component (the amplitude field), as indeed expected in the
Anderson-Higgs-Kibble mechanism, and there is also a (surviving)
massless component (the phase field) playing the r\^ole of the NG
mode. In the following we show that such a massless component is
crucially involved in the phase locking of the e.m. and matter
fields.

The emerging picture is then the following. The system may be
prepared with initial conditions dictated by the conservation of
the particle number and given by eqs. (\ref{4f}) and (\ref{4f2}),
where the value of the parameter $\theta_{0}$ is in principle
arbitrary within reasonable physical conditions. According to the
field equations (\ref{10}), the system then evolves towards the
minimum energy state where $|a_{0}(t)|^{2} \neq 0$ as in eq.
(\ref{29}) and the amplitude $|u(t)|^{2}$ departs from its initial
zero value. This implies  a succession of (quantum) phase
transitions \cite{DelGiudice:1988tc} from the initial $u_{0} = 0$
symmetric vacuum to the asymmetric vacuum $|u(t)|^{2}  \neq 0$,
which means that in eq. (\ref{20}) $\theta_{0}$ has to be greater
than $\frac{\pi}{4}$. In this way the lower bound for $\theta_{0}$
is dynamically fixed and the phase symmetry is dynamically broken
in the process of phase transition to the coherent regime. The
r\^ole of the  phason mode $\varphi$ is to recover such a
symmetry, thus re-establishing the gauge invariance of the theory.
This is done through the emergence of the coherence implied by the
phase locking between the matter field and the e.m. field. Let us
see how this happens.

\subsection{The phase locking}

As shown above, provided $\theta_{0}
> \frac{\pi}{4}$, a time-independent amplitude $U(t) \equiv
\overline{U}$ is  compatible with the system dynamics (e.g.  the
ground state value of $A_{0}$ in eq. (\ref{29}) implies
$\overline{U} = const.$, as noticed above). Eqs. (\ref{24c}) and
(\ref{24d})  with $\al_{m} \equiv \al = \de_{1}(t) - \de_{0}(t) -
\varphi(t)$ are
\bea \lab{24a}  \dot{U}(t) &=& 2\Om A_{0}(t)A_{1}(t) \cos \al (t) ~,\\
\lab{24b} \dot{\varphi}(t) &=& 2\Om
\frac{A_{0}(t)A_{1}(t)}{U(t)}\sin \al (t) ~. \eea
We thus see that $\dot{U}(t) = 0$, i.e. a time-independent
amplitude $\overline{U} = const.$ exists, if and only if the phase
locking relation
\be \lab{25} \al = \de_{1}(t) - \de_{0}(t) - \varphi (t) =
\frac{\pi}{2} \ee
holds. In such a case,
\be \lab{26}  \dot{\varphi}(t) =  \dot{\de_{1}}(t) -
\dot{\de_{0}}(t) = \om ~.\ee
which shows that any change in time of the difference between the
phases of the amplitudes $a_{1}(t)$ and $a_{0}(t)$  is compensated
by the change of the phase of the e.m. field. When eq. (\ref{25})
holds we also have $\dot{A_{0}} = 0 = \dot{A_{1}}$ (cf. eqs.
(\ref{24aa}), (\ref{24ab})). Provided $\theta_{0} > \frac{\pi}{4}$,
the phase relation (\ref{25}) can be thus regarded as a further
constant of motion implied by the dynamics: $\dot{\al} = 0$. It
expresses nothing but the gauge invariance of the theory. Since
$\de_{0}$ and $\varphi$ are the NG modes, eqs. (\ref{25}) and
(\ref{26}) also exhibit the coherent feature of the collective
dynamical regime: the system of $N$ dipoles and of the e.m. field is
characterized by the "in phase" dynamics expressed by eq. (\ref{25})
(phase locking). In other words, the gauge invariance of the theory
is preserved by the dynamical emergence of the coherence between the
matter field and the e.m. field. In such a regime we have
\be \lab{26g} \overline{A}_{0}^{2} - \overline{A}_{1}^{2} =
\cos^{2} \theta_{0} - \frac{1}{3}\sin^{2} \theta_{0} + 2
\overline{U}^{2} \neq 0 ~, \ee
to be compared with $A_{0}^{2}(t) - A_{1}^{2}(t) \approx 0$ at the
thermal equilibrium in the absence of the collective dynamical
regime  discussed here. Eq. (\ref{26g}) shows the relevant r\^ole
played by the occurrence of a  time-independent e.m. amplitude
$\overline{U}^{2}$; the collective dynamical regime, which sets in
for $\theta_{0} > \frac{\pi}{4}$, allows that a non-zero permanent
polarization $P_{{\bf n}}$ appears when  an electrical field is
applied, as discussed in deriving eq. (\ref{4pb}). In the
following we will come back on this point.

In conclusion we  recognize that, starting at $t = 0$ from the
initial condition $|u(0)|^{2} = 0$, and, correspondingly, from the
zero order parameter $u_{0} = 0$,  a non-zero time--independent
e.m. amplitude can develop (phase transition), provided
$\theta_{0} > \frac{\pi}{4}$, as an effect of the radiative
dipole-dipole interaction. This results in turn in the phase
locking (\ref{25}) and in the subsequent coherence in the time
behavior of the phase fields (cf. eq. (\ref{26})). Eqs. (\ref{25})
and (\ref{26}) show the r\^ole played by the phason field
$\varphi$ in recovering the gauge invariance in the process of
phase transition to the collective dynamical regime.

In the collective dynamical regime  considered above the values of
the amplitudes $A_{0}$ and $A_{1}$ are related to the  amplitude
$\overline{U}$ through the relations (\ref{4b}) and (\ref{16a}).
Moreover, we also obtain
\be \lab{27a}  A_{0}^{2}  = \frac{1}{3} [1 + \cos^{2} \theta_{0} +
(1 - \frac{1}{4} \sin^{2} 2 \theta_{0})^{1/2}]  ~. \ee
which used in (\ref{27}) shows that the oscillations around the
ground state for $A_{0}$ have pulsation $\nu = 2 \sqrt{2}(1 -
\frac{1}{4} \sin^{2} 2 \theta_{0})^{1/4}$.

The physical meaning of the phase locking can be stated as follows.
The gauge arbitrariness of the field $A_{\mu}$ is meant to
compensate exactly the arbitrariness of the phase of the matter
field in the covariant derivative $D_{\mu} = \pa_{\mu} - i
gA_{\mu}$. Should one of the two arbitrarinesses be removed by the
dynamics, the invariance of the theory requires the other
arbitrariness, too,  must be simultaneously removed, namely the
appearance of a well defined phase of the matter field implies that
a specific gauge function  must be selected. The above link between
the phase of the matter field and the gauge of $A_{\mu}$ is stated
by the equation $A_{\mu} = \pa_{\mu} \varphi$ ($A_{\mu}$ is a pure
gauge field). When $\varphi({\bf x},t)$ is a regular (continuous
differentiable) function then ${\bf E}=-\frac{\pa {\bf A}}{\pa t} +
{\bf \nabla} A_{0}= (-\frac{\pa }{\pa t}{\bf \nabla} +{\bf \nabla}
\frac{\pa }{\pa t})\varphi  = 0$, since in such a case time
derivative and the gradient operator can be interchanged.
Analogously, in the space of the regular functions $\varphi({\bf
x},t)$ it is ${\bf B}={\bf \nabla} \times {\bf A} = {\bf \nabla}
\times {\bf \nabla} \varphi= 0$. Thus the existence of nonvanishing
fields {\bf E} and  {\bf B} in a coherent region implies that the
time and space derivatives should act on a space larger than the
space of regular functions, namely $\varphi({\bf x},t)$ should
exhibit a (divergence or a topological) singularity within the
region \cite{Alfinito:2001mm}. This is precisely what is observed,
e.g., in type II superconductors when penetrated by  the lines of a
quantized flux in a vortex core.

\section{Discussion and conclusive remarks}

We have seen above that the rescaling of the field by the factor
$\sqrt{N}$ (cf. eq.(\ref{4})) induces the enhancement by the same
factor of the coupling constants appearing  in the field equations
(\ref{10}) (see also the coupling $\Om$ introduced in eqs.
(\ref{11a})-(\ref{11c})).  This implies that for large $N$ the
collective interaction time scale is much shorter (by the factor
$\frac{1}{\sqrt{N}}$) than the short range interactions among the
atoms. Hence the mesoscopic/macroscopic stability of the system vs
the quantum fluctuations in the short range dynamics of the
microscopic components. For the same reason, for sufficiently large
$N$ the collective interaction is protected against thermal
fluctuations. Indeed, thermal fluctuations could affect the
collective process only when $kT$ is comparable or larger than the
energy gap, whose value thus determines the height of the
protection; the larger the energy gap, more robust the protection.
We do not present in this paper an estimate of the energy gap. We
will do that in a future work. The r\^ole of the factor $\sqrt{N}$
in setting the time scale of the system can be understood also in
connection with the influence on the system of atoms of an electric
field ${\bf E}$ due, e.g., to an impurity (or to any other external
agent). By closely following ref. \cite{almut}, we will see indeed
that for large $N$ the system of atoms behaves as a collective
whole. The interaction ${\cal H} = - {\bf d} \cdot {\bf E}$ (eq.
(\ref{4p})) with the electrical field can be written \cite{Knight}
as
\be \lab{27b} H = \hbar \ga (b^{\dag} \sigma^{-} + ~b
\sigma^{+})~, \ee
which is a Jaynes-Cummings-like Hamiltonian, as already mentioned in
Section I in connection with Haken analysis. In eq. (\ref{27b})
$\ga$ is the coupling constant which is proportional to the matrix
element of the atomic dipole moment  and to the inverse of the
volume square root $V^{-1/2}$, $b$ is the electric field quantum
operator, $\sigma^{\pm}$ are the atomic polarization operators. Let
$|0 \rangle_i$ and  $|1 \rangle_i$, $i =1,...,N$, denote the ground
state and the excited state of each of the $N$ two-level atoms,
respectively, associated to the eigenvalues $\mp {1 \over 2}$ of the
operator $\sigma_{3i} = {1 \over 2} ( |1 \rangle_{ii} \langle 1| -
|0 \rangle_{ii} \langle 0|)$, no summation on $i$ (see e.g. Section
III.6 of ref. \cite{HakenBook} for the formal equivalence of the
system of two-level atoms with a system of $\frac{1}{2}$ spins). The
operators $\sigma_i^+= |1 \rangle_{ii} \langle 0|$ and $\sigma_i^-=
(\sigma_i^+)^\dagger$ generate the transitions between the two
levels induced by the action of the electric field.   The $N$-atom
system is thus described by $\sigma^\pm = \sum_{i=1}^N \sigma_i^\pm
, ~ \sigma_3 = \sum_{i=1}^N \sigma_{3i}$ with the fermion-like su(2)
algebra
\begin{equation}
\label{su2} [\sigma_3, \sigma^\pm] = \pm \sigma^\pm \, , ~~~
[\sigma^-,\sigma^+] = - 2 \sigma_3 \, .
\end{equation}
Suppose that the electric field action induces the transition $|0
\rangle_i \rightarrow |1 \rangle_i$ for a certain number of atoms,
say $l$ (as far as $N \gg l$ our conclusions will not be affected
by the fact that initially some of the atoms  are not in their
ground state). The system state may be then represented as the
normalized superposition  $|l \rangle$ given by
\begin{eqnarray} \label{L}
|l \rangle &\equiv& \big( |0\rangle_1 |0\rangle_2
...|0\rangle_{N-l}
|1\rangle_{N-l+1} |1\rangle_{N-l+2} ...|1 \rangle_N + \, . \, . \, . \nonumber \\
&& + |1\rangle_1 |1\rangle_2 ... |1\rangle_l |0\rangle_{l+1}
|0\rangle_{l+2} ... |0 \rangle_N \, \big) / {\textstyle \sqrt{N
\choose l}} \, . ~~~
\end{eqnarray}
The difference between the number of atoms in the excited state
and the ones in the ground state is measured by $\sigma_3$:
\be \lab{re2} \langle l| \sigma_3 |l \rangle = l - {1 \over 2}N ~
\ee
and the non-zero value of this quantity (proportional to the
system polarization) signals that the dipole rotational ($SU(2)$)
symmetry is broken. Operating with $\sigma^{\pm}$ on $|l \rangle$
gives:
\bea \nonumber \sigma^+ |l \rangle  &=& \sqrt{l+1} \sqrt{N-l} |l+1
\rangle ~,\\
\sigma^- |l \rangle  &=& \sqrt{N- (l-1)} \sqrt{l} ~|l-1 \rangle ~.
\label{rel} \eea
Eqs.~ (\ref{re2}) and (\ref{rel}) show that $\sigma_3$ and
$\frac{\sigma^\pm}{\sqrt{N}}$  are represented on $|l \rangle $ by
\bea \nonumber \sigma_3 &=& S^+ S^- - {\textstyle {1\over 2}}
N~,\\
\frac{\sigma^+}{\sqrt{N}} &=&  S^+ \sqrt{1 - {S^+ S^- \over N}} ,
~~~\frac{\sigma^-}{\sqrt{N}} = \sqrt{1 - {S^+ S^- \over N}} S^- ,
 \label{sigma}
\eea
where $S^- = (S^+)^\dagger$, $[S^-, S^+] = 1$, $S^+  |l \rangle  =
\sqrt{l+1} |l+1 \rangle $ and $S^- |l \rangle  = \sqrt{l} ~ |l-1
\rangle $, for any $l$. Eqs.~(\ref{sigma})  are the
Holstein-Primakoff non-linear boson realization of $SU(2)$
\cite{Holstein:1940zp,Shah:1974ha}. $\frac{\sigma^\pm}{\sqrt{N}}$ in
eqs.~(\ref{sigma})  still satisfy the su(2) algebra (\ref{su2}).
However, since for $N \gg l$  eqs.~(\ref{rel}) give
\begin{equation} \label{relc}
\frac{\sigma^\pm}{\sqrt{N}} \, |l \rangle  =   S^\pm \, |l \rangle
~,
\end{equation}
the su(2) algebra (\ref{su2}) contracts in the large $N$ limit to
the (projective) e(2)  algebra (or Weyl-Heisenberg algebra)
\cite{Wigner,DeConcini:1976uk,almut}
\begin{equation} \label{e2}
[ S_3,S^\pm] = \pm S^\pm \, , ~~~ [S^-, S^+] = 1 \, .
\end{equation}
where $S_{3} \equiv \sigma_{3}$. From eqs.~(\ref{relc}) and
(\ref{e2}) we see that, for large $N$, $S^\pm$ denote the creation
and annihilation {\it boson} operators associated to the quanta of
collective dipole waves excited by the electric field. The
interaction (\ref{27b}) can be now written in terms of $S^\pm$ as
\be \lab{27b1} H = \hbar \sqrt{N}\ga (b^{\dag} S^{-} + ~b S^{+})~,
\ee
We thus conclude that in the large $N$ limit the collection of
single two-level (fermion-like) atoms appears as a collective
bosonic system. The original coupling of the individual atoms to
the field gets enhanced by the factor $\sqrt{N}$ and appears as
the coupling of the collective modes $S^{\pm}$ (the system as a
whole) to the field. We observe that, as shown by eq. (\ref{4pb}),
the polarization persists as far as $\tau$ is non-zero, namely as
far as the field ${\bf E}$ is active (i.e. ${\cal H} \neq 0)$. The
system finite size prevents indeed from having a persistent
polarization surviving the ${\cal H} \rightarrow 0$ limit
\cite{Alfinito:2001mm}. In such a limit the dipole rotational
symmetry is thus restored.

Finally, we note that the collective dynamical features presented
here are not substantially affected by energy losses from the system
volume, which  we have not considered in the discussion above. These
losses are related with the different lifetimes of our different
modes, according to the different time scales associated to the
pulsations $m_{0}$, $\sigma$ and $\mu$. An analysis of energy losses
when the system is enclosed in a cavity has been presented elsewhere
in connection with the problem of efficient cooling of an ensemble
of $N$ atoms \cite{almut}. Another problem which we have not
considered in this paper is the one related to how much time the
system demands to set up the collective regime. This problem, which
is a central one in the domain formation in the Kibble-Zurek
scenario, will be the object of our study in a future work. Here we
remark only that, since the correlation among the elementary
constituents is kept by a pure gauge field, the communication among
them travels at the phase velocity of the gauge field.

$$         $$
\noindent {\bf Acknowledments} The authors are grateful to Prof. H.
Haken and to Prof. T. W. Kibble for useful discussions and suggestions. We thank
also Dr. A. De Ninno and Dr. M. Pettini for comments. Partial
financial support from Miur, INFN and the ESF Program COSLAB is
acknowledged.


\end{document}